\documentclass[prd,twocolumn,tightenlines,eqsecnum,nofootinbib,amsmath,amssymb,showpacs]{revtex4}

\input epsf
\usepackage{graphicx}

\begin{document}

\newcommand{\la}{\langle}
\newcommand{\ra}{\rangle}
\newcommand{\Tr}{{\rm Tr}}
\renewcommand{\Im}{{\rm Im}\,}
\newcommand{\be}{\begin{equation}}
\newcommand{\ee}{\end{equation}}
\newcommand{\bea}{\begin{eqnarray}}
\newcommand{\eea}{\end{eqnarray}}
\newcommand{\nl}{\nonumber \\}
\newcommand{\ket}[1]{|#1\ra} 
\newcommand{\sep}{\hspace{0,4cm},\hspace{0,4cm}}
\newcommand{\DD}{{\cal D}}
\newcommand{\psibar}{\,\overline{\phantom{I}}\!\!\!\!\psi}
\newcommand{\Dslash}{\,/\!\!\!\!D}
\newcommand{\Kslash}{\,/\!\!\!\!K}
\newcommand{\Lslash}{\,/\!\!\!\!L}
\newcommand{\Pslash}{\,/\!\!\!\!P}
\newcommand{\Qslash}{\,/\!\!\!\!Q}
\newcommand{\pslash}{\,/\!\!\!p}
\newcommand{\qslash}{\,/\!\!\!q}
\newcommand{\Pislash}{\,/\!\!\!\!\Pi}
\newcommand{\Dlr}{\stackrel{{}_{\,\leftrightarrow}}{D}}
\newcommand{\sgn}{{\mbox{sgn}}}
\newcommand{\Eq}[1]{{\rm Eq.~}(\ref{#1})}
\newcommand{\eq}[1]{{\rm Eq.~}(\ref{#1})}
\newcommand{\eqs}[1]{{\rm Eqs.~}(\ref{#1})}
\newcommand{\refeq}[1]{(\ref{#1})}
\newcommand{\OO}{{\cal O}}
\newcommand{\QCD}{{\mbox{QCD}}}
\newcommand{\gsim}{\stackrel{>}{{}_{\sim}}}
\newcommand{\lsim}{\stackrel{<}{{}_{\sim}}}
\newcommand{\EM}{{\rm EM}}
\newcommand{\rmO}{{\mathcal O}}
\newcommand{\rmOm}{{\mathcal O}_m}
\newcommand{\alphas}{\alpha_{\textrm{s}}}
\newcommand{\dA}{d_{\textrm A}}
\newcommand{\dF}{d_{\textrm F}}

\newcommand{\nF}{n_{\textrm F}}
\newcommand{\CA}{C_{\textrm A}}
\newcommand{\CF}{C_{\textrm F}}
\newcommand{\TF}{T_{\textrm F}}
\newcommand{\vecq}{\vec{\,q}}
\newcommand{\mubar}{{\mu_{\overline{\textrm{MS}}}}}
\def\cs{c_\textrm{s}}
\def\EE{\mathcal{E}}
\def\PP{\mathcal{P}}

\title{Asymptotics of thermal spectral functions}
\author{S.~Caron-Huot}
\affiliation
    {%
    Department of Physics,
    McGill University, 
    3600 University St.,
    Montr\'{e}al, QC H3A 2T8, Canada
    }%
\date {March 2009}

\begin {abstract}%
    {%
We use operator product expansion (OPE) techniques to study
the spectral functions of currents and stress tensors
at finite temperature,
in the high-energy time-like region $\omega\gg T$.
The leading corrections
to these spectral functions are
proportional to $\sim T^4$ expectation values in general,
and the leading corrections $\sim g^2T^4$ are calculated
at weak coupling, up to an undetermined coefficient in the shear
viscosity channel.
Spectral functions are shown to be infrared safe,
in the deeply virtual regime, up to order $g^8T^4$.
The convergence of (vacuum subtracted)
sum rules in the shear and bulk viscosity
channels is established in QCD to all orders in perturbation theory,
though numerically significant tails $\sim T^4/(\log\omega)^3$
are shown to exist in the bulk viscosity channel.
We argue that the spectral functions of currents and stress tensors
in infinitely coupled $\mathcal{N}=4$ super Yang-Mills
do not receive any medium-dependent power correction.
  }%
\end {abstract}

\pacs{11.10.Wx, 11.10.Jj, 12.38.Mh}  

\maketitle

\section{Introduction}

This paper is devoted to applying
operator product expansion (OPE) \cite{ope}
techniques
to study asymptotics of real-time spectral
functions $\rho(\omega,\vecq)$ at finite temperature.
We will obtain the leading corrections
in an expansion in $T/\omega$ at large frequencies $\omega$,
with $T$ a characteristic
energy scale of the medium, to the spectral functions
of currents ($J^\mu$), scalar operators ($m\psibar\psi$) and stress tensors
(in the shear and bulk viscosity channels).

The zero frequency, zero momentum limits of spectral functions
are related in general
to hydrodynamical transport coefficients, which,
for the quark-gluon plasma as probed by the Relativistic Heavy Ion
Collider \cite{RHIC},
have been the focus of much recent work
(for a more ample discussion we refer the reader
to \cite{hydro} and references therein).
At finite time-like momenta, the vector channel spectral function
is related to the production rate of lepton pairs by the plasma
\cite{dilepton}.

The OPE techniques employed in this paper, on the other hand,
give information
on the deeply virtual time-like region, $\omega\gg T,|\vecq|$.
Spectral functions in this region probe the surrounding medium only
on short time and length scales,
which is why they are related, by the OPE, to the expectation values
of local operators.
The OPE analysis thus does not cover the region
$\omega\sim T$ where the thermal corrections to the shape of
$\rho$ are the most important and where most of the, e.g.,
lepton pair emission, occurs.

This paper will cover the following applications.
First of all, the compact expressions obtained in the OPE regime
can be used as simple consistency checks on more complete
calculations.
For example, we will prove that the leading thermal corrections
to spectral functions in QCD are proportional in general
to $T^4$,
as was observed in the early calculations \cite{dilepton},
and we will give their coefficients in certain instances.
Also, infrared divergences can be fully characterized in the OPE regime:
we will show that, beginning at order $g^8T^4$, but not before,
certain spectral functions cease to be computable perturbatively
due to the so-called Linde problem.

An interesting sum rule was recently proposed
by Kharzeev and Tuchin \cite{kharzeev} and
by Karsch, Kharzeev, and Tuchin \cite{karsch},
and used to estimate the QCD bulk viscosity near the deconfinement
phase transition.
We will show that their sum rule is sensitive to a numerically
important ultraviolet tail,
ignored in \cite{kharzeev} and \cite{karsch},
which could affect
their analysis away from very close to the phase transition.

In the shear viscosity channel we will demonstrate
the convergence of sum rules in asymptotically free theories.
Furthermore, we will argue that the left-hand side of such sum rules
is saturated by a one-loop calculation, which however
will not be performed here.
We will also observe the possibility that such sum rules could possess,
in certain theories, strong ultraviolet tails making them
discontinuous in the free theory limit ($g^2\to0$).

Finally we will
briefly study spectral functions in the strong coupling limit of
$\mathcal{N}=4$ super Yang-Mills.
We will argue that power corrections to spectral functions
at large virtuality are restricted to polynomial terms in the momenta
and forbidden for currents and stress tensors,
generalizing observations of Teaney \cite{teaney06}.

This paper is organized as follows.
After describing our formalism in section \ref{sec:formal},
we apply it in the weak coupling regime to the
spectral functions enumerated above, in section \ref{sec:app}.
The physical implications of these results are then discussed
in section \ref{sec:disc}.
Finally, in an Appendix \ref{app:calc}, we reproduce a two-loop
diagrammatic calculation which confirms the OPE prediction in the vector
channel.

\section{Formalism}
\label{sec:formal}

\def\GE{G_\textrm{E}}
\def\omegaE{\omega_\textrm{E}}

The physical basis of the (Euclidean) OPE \cite{ope}
is the separation of scales between that a short-sized
probe $\Delta x$ and that of a typical wavelength
$\sim T^{-1}$ in a medium, leading to useful asymptotic
expansions in $(T\Delta x)$.
In Fourier space for the expectation value of a two-point function
of operators $\OO_{1,2}$
this gives rise to asymptotic expansions
at high momenta,
suppressing arguments other than the frequency:
\be
  \GE{}_{12}(\omegaE)\sim
\sum_n \la \OO_n \ra \frac{c_{12}^n}{\omegaE^{d_n}}\,.
  \label{schemG}
\ee
The locality of the operators $\OO_n$ (which we will
assume to be Hermitian, without loss of generality)
follows from $T\ll \omega_E$.
The powers $d_n$ are determined by
renormalization group equations (RGE), to be reviewed shortly.

Naively taking twice the imaginary part
of the analytic continuation of \eq{schemG}
to Minkowski frequencies $\omega$
would yield asymptotics for the spectral function,
\be
  \rho_{12}(\omega) \sim \sum_n \la \OO_n\ra
 \,2\,\Im \left(\frac{c_{12}^n}{({-}i\omega)^{d_n}}\right).
\label{naivesum}
\ee
with $\rho(\omega)$ the commutator function
$\langle [\OO(\omega),\OO(x{=}0)] \rangle$.
\eq{naivesum} is ``naive'' in that the asymptotics of an analytic
function in one complex direction do not, in general,
determine its asymptotics along other directions.
It is nevertheless correct
whenever $\rho(\omega)$ does admit an asymptotic expansion
in inverse powers of $\omega$ (or logarithms),
as will be proved shortly.

\eq{naivesum} is the main result of this section.
We read it as a Minkowski-space version of the OPE:
heuristically, the locality of its operators is a consequence
of the small time during which a pair of high-energy
particles, created by a high-frequency operator,
can travel before it is reabsorbed
by its complex conjugate.
Due to this picture, once its coefficients are determined
we expect it to remain valid
even in out-of-equilibrium situations,
where an Euclidean formulation is not available.


\subsection{Dispersion relations}
\label{sec:disp}

Here we justify the passage from \eq{schemG} to
\eq{naivesum}, \emph{assuming} that $\rho(\omega)$ admits
an asymptotic expansion in inverse powers of $\omega$ and logarithms.
The argument is based on the
dispersive representation of the Euclidean correlator%
\footnote{
 At finite temperature, this gives a distinguished
 analytic continuation of $\GE$
 from the discrete set of Matsubara frequencies at which it is strictly
 defined \cite{kapustagale}.
 Substituting $\omegaE=-i\omega$ with $\omega$ in the upper-half plane,
 \eq{dispersion} always coincides with
 the retarded function $G_\textrm{R}(\omega)$.
},
\be
\GE(\omegaE)= P_n(\omegaE) +
\int_{-\infty}^\infty \frac{d\omega'}{2\pi(\omega'{-}i\omegaE)}
\rho(\omega')\,.
\label{dispersion}
\ee
Note that $\rho$ is always real (or a Hermitian matrix),
$\rho=2\Im G$.
$P_n(\omegaE)$ is a polynomial in $\omegaE$ that is not
determined by the spectral density.  In general
\eq{dispersion} is ultraviolet
divergent and a subtracted integral must be used,
but this does not interfere with the present argument.

The basic point is that,
if an upper cut-off $|\omega'|<\Lambda$ were imposed on
the frequency integration in \eq{dispersion},
the resulting $\GE(\omega)$ would admit an expansion
in purely integral powers of $1/\omega$ at large $\omega$.
Specifically, for
$\rho(\omega)$ bounded by $|\omega|^{-k-\epsilon}$  ($\epsilon>0$)
at large
$\omega$, the $k^{\rm th}$ derivative of \eq{dispersion}
with respect to $1/\omega$ is shown to vanish at $\omega=\infty$.
This shows that the asymptotic expansion of $\rho$
directly translates into
one for $\GE$ modulo integral terms (e.g., terms
that are killed by taking derivatives).

Thus it suffices to match the asymptotic expansions of
\eq{schemG}
and \eq{naivesum} termwise.
Two cases must be distinguished:
non-integral power terms (or powers times logarithms),
and purely integral powers.

The former case is dealt with with the dispersive transform of
a power law tail $\rho(\omega)\propto \omega^{-\alpha}$,
\be
\int_{\omega_0}^\infty \frac{d\omega'}{2\pi(\omega'{-}i\omegaE)}
\frac{1}{\omega^\alpha} \sim
\frac{1}{2\sin \pi \alpha} \frac{1}{(-i\omegaE)^\alpha} +
\sum_n \frac{D_n \mu^{n-\alpha}}{\omega^n},
\label{exdispersion}
\ee
where $\mu_0$ is some infrared cut-off and the sum
is an asymptotic series with $n$ integers.
The coefficients $D_n$ in this sum depend
on the details of
infrared data (here on $\omega_0$)
but the non-analytic term is a clean reflection of the
large $\omega$ behavior of $\rho$, as expected.
Its imaginary part at real $\omega$
produces the right asymptotics.  In particular,
the asymptotics at positive and
negative $\omega$ can be reconstructed independently
since their contributions are out of phase at real $\omegaE$.
Taking derivatives of \eq{exdispersion} with respect to $\alpha$ gives
identities for integrals with logarithms, for which the same argument
applies.

The case of purely integral powers in $\rho$, $d_n$ integer,
is special because
the Euclidean non-analyticity is a single logarithm:
it could cancel out between the contributions of positive and negative
$\rho$. Such cancellations would occur when $d_n$ is even and
its contribution to $\rho(\omega)$ is even in $\omega$,
or when $d_n$ is odd and its contribution is also odd.
In these cases,
the Euclidean function would be proportional to $1/(-i\omegaE)^n$
with an imaginary coefficient, e.g. it has a ``wrong'' phase.
On the other hand, integral terms coming from small frequencies
in \eq{dispersion}, are proportional to
$1/(-i\omegaE)^n$ with a real coefficient.  Thus the two are cleanly
separated by their phases, and we conclude that in all cases
the asymptotics of $\rho$ can be recovered from those of $\GE$.

In perturbation theory we do not expect such ``wrong phase''
contributions to the Euclidean OPE, and in any case certainly none
appears at the relatively low orders to which we will be working
in this paper.
Our corrections to spectral functions will come
solely from non-analytic terms in $\GE$.

We now comment the assumption that
$\rho(\omega)$ admits an expansion in inverse
powers of $\omega$, which we have assumed in proving
\eq{naivesum}.
Possible violations of it at the nonperturbative level
(due, for instance, to oscillating terms)
are discussed in \cite{shifman00}.
However, in perturbation theory we find it
hard to see how it could fail,
for instance
because there is no scale to provide an oscillation rate.
Thus this assumption will be made throughout this paper.

\subsection{Renormalization group equations}
\label{sec:RGE}

The OPE is based on a systematic separation of
infrared and ultraviolet contributions (an interesting discussion may be
found in \cite{novikovetal}).
A factorization scale $\mu$ (for us, $T\lsim \mu\ll \omega$)
is introduced, and all vacuum fluctuations
from above this scale are integrated over, leaving more infrared and
state-dependent fluctuations to be accounted for by the expectation values
of operators.
Since $\mu\ll \omega$ these operators can be taken to be local,
in a systematic gradient expansion.
The restriction to vacuum fluctuations ensures that this yields
operator relations,
that is, the OPE holds in any quantum state.

Omitting Lorentz and internal indices,
this yields expansions of the form,
\be
\OO_1^{(\mu)}(p) \OO_2^{(\mu)} (x{=}0) \sim
\sum_i C_{12}^i (p,\mu,\mubar,g,m)  \OO_i^{(\mu)}(x{=}0)\,,
\label{ope1}
\ee
with $m$ and $g$
standing for various intrinsic mass scales and couplings of the theory
and $\mubar$ its renormalization scale.
The renormalized operators $\OO_i^{(\mu)}$ obey the RGE,
with $\gamma$ a matrix of anomalous dimensions,
\be
0=\left[\mu\partial_\mu + \gamma \right]
 \OO_i^{(\mu)}\,,
\label{operun}
\ee
from which we deduce, in the case that $\OO^{(\mu)}_{1,2}$ are independent of
$\mu$ (as for currents, which we will exclusively study in this paper),
the RGE for the OPE coefficients:
\be
 \left[\mu\partial_\mu - \gamma^T \right]
   C_{12}^{i}= 0.
\label{ope2}
\ee

Assuming the absence of microscopic scales between the factorization
scale $\mu$ and $p$,
the coefficient functions can depend only on three
scales: the momentum $p$, the factorization scale $\mu$
and the renormalization scale $\mubar$ of the theory.
The $\mubar$ dependence is determined by a RGE
\be
 \left[\mubar\frac{\partial}{\partial_\mubar} +
  \beta(g)\frac{\partial}{\partial g} +\ldots \right]
 \OO_1(p) \OO_2 = \sum_i D^i_{12}(p) \OO_i.
 \label{inhom}
\ee
The important point for us will be that the coefficients $D^i_{12}$
can only be polynomials in the momenta $p$. This is because correlators
at unequal positions are RGE-invariant
and Fourier transforms can always be made insensitive to
the coincidence limit by taking sufficiently many derivatives with
respect to momenta.

Were there no right-hand side to \eq{inhom} we would conclude,
on dimensional grounds,
that all logarithms in the coefficient functions have to be of the form
$\log(p^2/\omega^2)$.
\eq{inhom} shows that terms which are polynomial in $p$
can also contain logarithms of $\mubar$.%
 \footnote{ Examples of such terms are the logarithms
   in the Green's functions of free theories.  Since
   they originate from ultraviolet divergences they
   do not depend on $\mu$, and their non-analytic
   behaviors $\sim \log(p^2/\mubar^2)$ produce the free theory
   power tails in the spectral functions.
 }


\subsection{RGE in Minkowski signature}
\label{sec:mink}

Since we interpret \eq{naivesum} as a Minkowski-space version
of the OPE,
it is worth specifying how we mean the RGE 
of operators directly in Minkowski space:
we mean it to be exactly what it is in the Euclidean OPE,
namely vacuum fluctuations should be integrated over but not
state-dependent ones.
Provided equivalent regulators are used, this will reproduce
the standard running of the operators in Euclidean space.%
 \footnote{
  Examples of equivalent regulators include
  a sharp momentum cut-off on $\vec{p}$ and dimensional regularization.
 }

Variations on this procedure can easily lead to difficulties
due to transport phenomena (e.g., nonlocal phenomena),
as may be illustrated by a concrete example: that of computing
the contribution of a fluctuation at scale $\sim gT$
(with $g=\sqrt{4\pi\alphas T}$)
to the expectation value
of a local operator in a weakly coupled quark-gluon plasma.
In this case, it would be natural to fully integrate out the
scale $T$, which produces, at the one loop level,
the Hard Thermal Loop effective theory \cite{htlpapers}.
It is nonlocal with support on
the classical trajectories of plasma particles.
The conclusion is that, employing any reasonable procedure,
fully integrating out the scales
$gT\lsim \mu\lsim T$ in real time
must convert local operators at the scale $T$ to
\emph{nonlocal} operators at the scale $gT$.
This does not happen, however,
when only vacuum fluctuations are integrated over%
 \footnote{
  A heuristic way to understand this fact is by analogy
  to the situation in the band theory of metals,
  in which completely filled bands do not conduct but
  only partially filled bands do:
  the field-theoretic vacuum is akin to a completely filled band.
 },
as was assumed in the preceding subsection.

\section{Asymptotics of spectral functions}
\label{sec:app}

At weak coupling, the OPE coefficients for spectral functions,
\eq{naivesum}, are products of
Euclidean OPE coefficients and anomalous dimensions,
which we now compute in turn.

\subsection{Conventions}

For concreteness and simplicity
we study the Euclidean Yang-Mills theory coupled
to $\nF$ Dirac fermions of the same mass $m$,
$ S_{\rm E}=
  \frac{F_{\mu\nu}F_{\mu\nu}}{4g^2}
  + \sum_i \,\psibar_i (\pslash -im)\psi_i
$,
with $p^\mu$ the covariant momentum.
Some important operators (in Euclidean notation) will be
\begin{subequations}
\begin{align}
T^{\mu\nu}_{g}
 &= \frac{1}{g^2}\left[
   G^{\mu\alpha}G^{\nu}{}_\alpha
   -\frac{\delta^{\mu\nu}}{4} G^2 \right],
\\
T^{\mu\nu}_{f}
 &= \sum_i \psibar_i \frac{-i\Dlr^\mu \gamma^\nu + 
   -i\Dlr^\nu\gamma^\mu}{4}\psi_i
- \mbox{[Trace part]}\,,
\\
\rmOm &= -im\sum_i\psibar_i\psi_i\,.
\end{align}
\end{subequations}
$T^{\mu\nu}_g + T^{\mu\nu}_f$ is the traceless part
of the full stress tensor.
We will also consider the trace of the stress tensor,
\be
   T^\mu{}_\mu=
     \frac{b_0 G^2}{32\pi^2}
     + \mbox{[fermion terms]},
\ee
where $b_0= (\frac{11}{3}\CA {-} \frac43\nF\TF)$ is the leading coefficient
of the $\beta$-function
($\beta(\alphas)\approx -b_0 \alphas^2/2\pi$).

\subsection{Euclidean OPE coefficients}

\begin{figure}
\begin{center}
\includegraphics[width=0.48\textwidth]{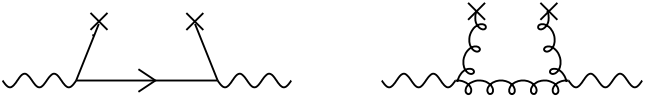}
\end{center}
\caption[OPE diagrams]{Feynman diagrams giving the leading
  order OPE coefficients for currents and stress tensors.}
\label{fig:ope}
\end{figure}

At the leading order in perturbation theory
the OPE of
currents $J^\mu=\sum_i\psibar_i\gamma^\mu\psi_i$
is given by the first diagram
of Fig.~\ref{fig:ope}.
A pedagogical introduction to this sort of calculation can be found
in \cite{shifmancalc}.
Up to dimension four operators we find:
\begin{align}
& J^\mu(q)J^\nu \! \sim\! \sum_i\psibar_i \!
  \left[
   \gamma^\mu  \frac{1}{\qslash{-}i\Dslash {-}im} \gamma^\nu
{-}  \gamma^\nu \frac{1}{\qslash{+}i\Dslash {+}im} \gamma^\mu
  \right]\!\!\psi_i
\nl
 \sim&
 \frac{2\epsilon^{\mu\nu\alpha\beta} q^\alpha}{q^2}
    \sum_i \psibar_i\gamma^\beta\gamma_5\psi_i
-\!\frac{2\left(\delta^{\mu\nu}{-}\frac{q^\mu q^\nu}{q^2}\right)}
     {q^2} \rmOm
\nl
 &+
  \frac{4}{q^4} T^{\alpha\beta}_{f}
   \left[
     \delta^{\mu\alpha}\delta^{\nu\beta} q^2
     {-}\delta^{\nu\beta} q^\mu q^\alpha
     {-}\delta^{\mu\beta} q^\nu q^\alpha
     {+}\delta^{\mu\nu} q^\alpha q^\beta
   \right],
\label{OPE1}
\end{align}
with $\gamma_5=\gamma_1\gamma_2\gamma_3\gamma_4$ and the covariant
derivative acting to its right on $\psi$.
We will drop the first term since it
does not contribute to spectral functions,
being scale-invariant.


In isotropic media we decompose
the current correlator into transverse and
longitudinal components: we let $q=(q^4,0,0,|\vecq|)$, set
$G^T= \la J^1 J^1\ra$, $G^L= \frac{q^2}{(q^0)^2}\la J^3 J^3\ra$
and employ $T^{ij}_{f,g}= -\frac13\delta^{ij}T^{44}_{f,g}$
for the traceless operators.
\Eq{OPE1} then gives:
\begin{subequations}
\label{OPE}
\bea
G^T(q) &\sim& \frac{8}{3q^4}\left(q_4^2-\vecq^2\right)
   T_f^{44}
  -\frac{2}{q^2}\rmOm\,,
\\
G^L(q) &\sim& \frac{8}{3q^2} T_f^{44}
  -\frac{2}{q^2}\rmOm \,.
\eea
Similarly, the OPE of $\rmOm$ (scalar channel) is,
\be
G^S(q) \sim \frac{3m^2}{q^2}\rmOm + \frac{4m^2}{q^4}
       q_{\alpha}q_{\beta} T_f^{\alpha\beta} +\ldots\,,
\ee
that of the trace anomaly $T^\mu_\mu$ (bulk viscosity
channel) is,
\be
G^{\zeta}(q) \sim
     b_0^2 \alphas^2
    \left( 
      4\frac{q_\mu q_\nu}{q^2} T^{\mu\nu}_g + \frac{G^2}{g^2}
    \right),
\label{opeZeta}
\ee
and that of the shear mode $T^{12}$ of the stress tensor,
assuming isotropy, is:
\be
 G^{\eta}(q) \sim 
    \frac{2}{3q^2}\left(q_4^2 - \vecq^2\right) T^{44}_g 
  + 
 \frac{1}{6} \frac{G^2}{g^2}
 \,.
 \label{opeEta}
\ee
\end{subequations}
The fermion contributions to \eqs{opeZeta}
and (\ref{opeEta}) begin at dimension-6
and have been dropped.
However, we must be kept in mind that the OPE, like any two-point
function, is really defined only modulo contact terms (terms purely
polynomial in momenta).

\subsection{Contact terms}
\label{sec:contact}

The OPE coefficients of $G^2$ in \eqs{opeEta} and (\ref{opeZeta})
are purely polynomial in $q$.
According to the discussion at the end of section \ref{sec:RGE},
this means we have to decide whether the operators get
evaluated at the scale $\mubar$ or $\omega$; alternatively,
contact terms depending only on $\mubar$
could be freely shifted in and out of the OPE as just mentioned.

In the shear channel \eq{opeEta} it turns out that, had we computed
the full OPE coefficient for general $T^{\mu\nu}T^{\alpha\beta}$,
we would have found that its coefficient is purely
polynomial and non-transverse (e.g., leading to 
$q_\mu T^{\mu\nu} T^{\alpha\beta} \neq 0$).
Evaluating this operator at the scale $\omega$
would lead to a non-transverse spectral function, which is impossible.
Therefore, the $G^2$ term in \eq{opeEta} must be a pure contact term that
runs with the scale $\mubar$ and does not contribute to spectral functions.
This issue is discussed in \cite{novikovG}.

A similar ambiguity makes it possible to shift the coefficients
of $T^{44}_g$ and $T^{44}_f$ in \eq{opeEta} by $p$-independent
constants.  Although we believe that this could, in principle, be settled
by studying the Ward identities obeyed by the full OPE of
$T^{\mu\nu} T^{\alpha\beta}$, as in the above paragraph, this will
not be done in this paper.  This will translate in an indeterminacy
for our shear channel spectral function.

The Ward identities are much harder to exploit in the bulk channel
because the trace $T^\mu{}_\mu$ is subleading at weak coupling.
Thus it seems hard to determine,
without an explicit calculation of running coupling effects,
at which scale $g^2G^2$ is to be evaluated in this channel.
Such a calculation will not be attempted in this work,
so we will only be able to determine the asymptotics in this channel
modulo $g^2G^2$.
The term proportional to $q_\mu q_\nu T^{\mu\nu}_g/q^2$
in \eq{opeZeta} is
unambigous, however, since the Lorentz covariance of the OPE
forbids the addition of spin-2 contact terms to spin-0
operator products.

\subsection{Anomalous dimensions}

The anomalous dimensions matrix
of dimension-four, spin-two operators reads \cite{peskin},
acting on the basis $(T^{\mu\nu}_g,T^{\mu\nu}_f)^T$,
\be
 \gamma = 
 \frac{\alphas}{3 \pi}
   \left(\begin{array}{ll}
      2\nF\TF &-4\CF \\
     -2\nF\TF & 4\CF
    \end{array} \right) +\rmO(\alphas^2),  \label{anom}
\ee
giving, e.g., the $\mu$ dependence of
$T^{\mu\nu}_f$ in \eq{OPE}:
\be
T^{\mu\nu}_f(\mu) \sim T^{\mu\nu}_f(\mu_0)
 + \frac{\alphas \log \frac{\mu_0^2}{\mu^2}}{3\pi}
  \left[2\CF T^{\mu\nu}_f {-} \nF\TF T^{\mu\nu}_g \right]
\ee
up to terms of order $\alphas^2$.

The combination $\rmOm$ does not run in perturbation
theory,
but the bare mass $m$ has $\mu$-dependence given by
$\gamma_m=3\alphas\CF/2\pi +\rmO(\alphas^2)$,
which must be included wherever it appears explicitly.  

In theories with massless quarks, $G^2$ does not run
at one-loop (at this order $G^2{\sim}T^\mu{}_\mu$, which does not run
to any order),
though explicit powers of $\alphas$ do run according to
$\gamma_{\alphas}=-\beta(\alphas)/\alphas=b_0\alphas/2\pi$.

\subsection{Spectral functions}

Frequency-dependent logarithms in $\GE$
are determined by the RGE of the operators entering
the right-hand side of the OPE, \eqs{OPE},
$\log (1/\mu^2)\to \log (1/\omegaE^2)$.
To obtain $\rho$ we take twice the imaginary part
at real $\omega$ (or employ \eq{naivesum}),
$\log (1/\omegaE^2)\to 2\pi$.
To help clarify our conventions we
recall the leading order, massless, zero-temperature results,
with
 $
  q^2=q_0^2{-}\vec{q}^2
 $:
$
 \left.\rho^T(q)\right|^\textrm{vac} =
 \left.\rho^L(q)\right|^\textrm{vac} =
  \frac{\nF\dF q^2}{6\pi}
$,
$
 \left.\frac{\rho^S(q)}{m^2}\right|^\textrm{vac} =
   \frac{\nF\dF q^2}{4\pi}
$,
$
 \left. \rho^\zeta(q) \right|^\textrm{vac} =
 \frac{b_0^2 \alphas^2 \dA q^4}{128\pi^3}
$,
$
 \left. \rho^\eta(q) \right|^\textrm{vac} =
  \frac{q^4}{80\pi} \left[ \dA + \frac12 \nF\dF\right]
$ \cite{teaney06}.
This way we find the leading (dimension-four) thermal corrections:
\begin{subequations}
\label{operesult}
\bea
\delta \rho^T(q)  &\sim&
\frac{16\alphas}{9q^2} 
 \frac{q_0^2{+}\vecq^2}{q^2}  
      \left[2\CF T^{00}_f {-} \nF\TF T^{00}_g \right],
 \label{Resa}
\\
\delta \rho^L(q)  &\sim&
 \frac{16\alphas}{9q^2}
      \left[2\CF T^{00}_f {-} \nF\TF T^{00}_g \right],
 \label{Resb}
\\
\delta \rho^S(q) &\sim&
  \frac{8\alphas m^2}{3q^2} \frac{q_\mu q_\nu}{q^2}
      \left[\frac{13}{2}\CF T^{\mu\nu}_f {-} \nF\TF T^{\mu\nu}_g \right]
\nl
&&
-\frac{9\alphas m^2\CF}{q^2} \rmOm,
\label{opescal}
\\
\delta \rho^\zeta(q)  &\sim&
  \frac{b_0^2 \alphas^3}{6\pi^2} \frac{q_\mu q_\nu}{q^2}
      \left[2\CF T^{\mu\nu}_f {-}
  (\nF \TF{+}\frac32 b_0) T^{\mu\nu}_g \right] \nl
&&
- C\frac{b_0^2\alphas^2}{4\pi} T^\mu{}_\mu,
 \label{opezeta}
\\
\delta \rho^\eta(q)  &\sim&
 \frac{4\alphas}{9}\!
\left(\!D{+}\frac{2\vecq^2}{q^2}\!\right)\!  
      \left[2\CF T^{00}_f {-} \nF\TF T^{00}_g \right]\!.
\label{opeeta}
\eea
\end{subequations}
The presently undetermined coefficients $C$ and $D$ are due
to the ambiguities discussed in subsection \ref{sec:contact},
with $C{=}D{=}1$ corresponding to choosing the
renormalization scale $\omega$ in \eqs{opeZeta} and (\ref{opeEta}).
$C$ and $D$ are in principle both
computable by more accurate calculations.
For simplicity, \eq{opezeta} accounts for the running of $G^2$
only for massless fermions, in general this operator can mix with $\rmOm$.
\eqs{opescal} and (\refeq{opezeta}) do not assume isotropy
and the anisotropic case of \eqs{Resa} and (\ref{Resb})
may be obtained starting from \eq{OPE1}.
We note that $\delta \rho^\eta=0$ when $\nF=0$.

The operators in \eq{operesult} are all Minkowski-signature
operators, with the energy density being $\EE=T^{00}={-}T^{44}$.
Please also note that \eqs{operesult} are written in $({+}{-}{-}{-})$
metric convention,
so both $q^2$ and $T^\mu{}_\mu$ have opposite sign relative to there
Euclidean cousins (e.g., $T^\mu{}_\mu=\EE{-}3\PP$ in \eq{operesult}).
In Minkowski space with fermionic action
$\sum_i \psibar_i (\pslash-m)\psi_i$, $\rmOm=m\sum_i\psibar_i\psi_i$.

At the Stefan-Boltzmann (free) level,
\begin{subequations}
\label{loexp}
\bea
 \la T^{00}_g\ra &=& \frac{\pi^2T^4}{15}\dA,  \label{loexp1} \\
 \la T^{00}_f\ra &=& \frac{7\pi^2T^4}{60}\nF\dF, \\
 \la \rmOm \ra &=& \frac{m^2T^2}{3}\nF\dF +\rmO(m^4).
\eea
\end{subequations}

\eqs{operesult}, together with (\ref{loexp}), constitute the main results
of this paper.


\section{Discussion}
\label{sec:disc}

\subsection{The photon spectral function}
\label{sec:photon}

One motivation for doing the present calculations
was to resolve, in a logically independent way,
a discrepancy in the literature regarding the asymptotics of
$\delta \rho^T(E)$.
This (in fact the complete spectral function for $\omega\gg gT$
at zero spatial momentum)
has been computed a long time ago
\cite{dilepton,gabellini} to order $\alpha_s$ (two-loop order),
and it was observed that the correction
was proportional to $g^2T^4/\omega^2$ at large energies $\omega$.
In contrast, a more recent calculation \cite{majumdergale}
instead observed a $g^2T^2$ behavior.

The OPE analysis presented in this paper makes it clear that
the dominant thermal effects must scale like $g^2T^4/\omega^2$, since the
lowest dimension of a (gauge-invariant) local operator
with a nontrivial anomalous dimensions in QCD is 4.
In particular, a $g^2T^2$ asymptotic behavior is forbidden
by the absence of local dimension-2 operators.
Thus we can confirm (at least qualitatively)
the early findings \cite{dilepton,gabellini}.
It is difficult here, however,
because of the different techniques employed,
to comment explicitly on the careful calculation
of \cite{majumdergale}.%
 \footnote{
  We do nevertheless agree with the main conclusion
  of \cite{majumdergale}, which was to rule out the
  infrared divergences claimed
  in \cite{kapusta} (see also the reply \cite{bigreply}).
 }

In Appendix \ref{app:calc}, we reproduce
the OPE result \eq{vectorres}
by means of a more standard diagrammatic
calculation in real-time perturbation theory.

For $\rho^T$ at $\vecq=0$ we find (upon
reinstating the well-known $T=0$ result, not computed here):
\be
 \rho^T(\omega) \!\sim \!\!\frac{\nF\dF \omega^2}{6\pi}
\!\!\left[1{+}\frac{3\alphas\CF}{4\pi}
  {+} \frac{16\pi^3\alphas\CF}{9}
     \frac{T^4}{\omega^4}
  {+}\OO(\alphas^2,T^6)
 \right]\!.
 \label{vectorres}
\ee
Interestingly, the correction, though parametrically small $\sim T^4/\omega^4$
at high energies,
has a large numerical prefactor, suggesting that
it could be useful down to not-so-large frequencies $\omega\gsim T$.
Comparison with the complete two-loop calculation
\cite{dilepton,gabellini}
should allow a precise determination of the regime of applicability
of the OPE, which will not be pursued here.

\subsection{Massive fermions ($m\gg T$)}
\label{sec:mikko}

Spectral functions of massive particles (with $m\gg T$)
at $\vecq=0$ have been considered in
\cite{mikko}.
In this case the fermionic condensates $T_f^{\mu\nu}$
and $\rmOm$ do not contribute,
and our results \eqs{operesult} for the thermal corrections
(e.g., ignoring $\rmO(\alphas)$ vacuum corrections) become
$
 \rho^T(q) \sim \frac{\dF q^2}{6\pi}
   \left[1  - \frac{32\pi^3\alphas\CF}{45}
     \frac{T^4}{q^4}+\ldots  \right]
$
and
$
 \rho^S(q) \sim \frac{\dF m^2 q^2}{4\pi}
 \left[1  - \frac{32\pi^3\alphas\CF}{45}\frac{T^4}{q^4}
   +\ldots  \right]
$,
in complete agreement with the findings%
 \footnote{
  To extract the complete asymptotics of \cite{mikko}
  we observe that the mass subtraction (4.3) must be undone from their
  final results (4.7) and (C.11).
  In their notation, at order $1/\omega^2$ this means adding
  $\frac{g^2\dF\CF M^2T^2}{2\pi \omega^2}$ to (4.7) and
  $\frac{g^2\dF\CF M^2}{8\pi}(-T^2+2M^2T^2/\omega^2)$ to (C.11).
 }
 of \cite{mikko}.

\subsection{Infrared divergences}
\label{sec:IR}

The OPE analysis completely determines the cancellation pattern of
infrared
divergences at high energies: the OPE coefficients contain
only infrared safe zero-temperature physics.
Infrared sensitivity can enter only through the
expectation values of the local operators.

The leading order corrections to spectral functions $\rho$
are proportional to $g^2$ anomalous dimensions
times tree-level expectation values of dimension-four operators.
The perturbative series
for such expectation values should be similar to that for the
thermodynamic pressure \cite{pressure}.
This implies that sensitivity of $\rho$ to the $gT$ scale
(and the necessity for Hard Thermal Loop resummation)
will first enter at order $g^5T^4/\omega^2$,
and that
nonperturbative physics associated with the $g^2T$ scale (the so-called
Linde problem \cite{linde})
will begin to contribute at order $g^8T^4/\omega^2$.

This situation should be contrasted with that for
the shape of the spectral
function at $\omega\lsim T$, for which infrared sensitivity
shows up much earlier.  At $\omega\sim T$, $gT$-scale physics
appears at order $g^3$ \cite{aurenchecarrington},
whereas at $\omega\sim gT$ it is important already at the $\OO(1)$
level \cite{braatensoft}.
Various integrals over the spectral functions, in the spirit
of the famous $T=0$ sum rules \cite{SVZ}, however, are
still governed by the Euclidean OPE, which
becomes sensitive to the $gT$ scale only at order $g^3$
and to the $g^2T$ scale at order $g^6$.
This seems to constrain, to a large extent,
the corrections to the shape of $\rho$ to only
move things around in frequency space.

\subsection{Bulk viscosity sum rules}
\label{sec:bulk}

A priori, the $\OO(1)$ power tails in $G^\zeta$, \eq{opezeta},
might appear sufficiently strong to make
even the difference
$\delta\GE^\zeta=\GE^\zeta{}{-}\GE^\zeta{}^\textrm{vac}$,
between the thermal Euclidean correlator of $T^\mu{}_\mu$ and its vacuum
limit, require a subtracted dispersive representation.
However, for asymptotically free theories,
the RGE invariance of $G^\zeta$ forces us to evaluate
the factors of $g^2$ in \eq{opezeta} at the scale $\omega$,
in which case $\rho(\omega)\sim 1/\log^2\omega$ at worse
and an un-subtracted dispersion relation for $\delta\GE(0)$ converges:
\be
 \delta\GE^\zeta(0) = \int_{-\infty}^\infty
  \frac{d\omega'}{2\pi \omega'}
 \delta\rho^\zeta(\omega').
 \label{dispLET}
\ee
Higher order corrections to OPE coefficients will be suppressed by
powers of $g^2(\omega')\sim 1/\log \omega'$
and will not affect convergence.
We are not making any assumption here
about the value of the coupling constant at the scale $T$, only
the scale $\omega'$ is important to the OPE coefficients.

On the other hand, in \cite{LET} the Euclidean correlator
$\delta \tilde{G}_\textrm{E}^\zeta(0)=\lim_{q\to 0} \lim_{\omega\to 0}
\delta G_\textrm{R}(\omega,q)$
is evaluated, by means of broken scale invariance
Ward identities (``low energy theorems'') \cite{LET},
\bea
\delta \tilde{G}_\textrm{E}^\zeta(0)
 &=& \left(T\frac{\partial}{\partial T} -4\right) (\EE-3\PP)
\nl
 &=& (\EE + \PP)\left(\frac{1}{\cs^2}-3\right) -4(\EE-3\PP),
\label{LET}
\eea
with $\EE=T^{00}$ and $\PP=T^{11}$.

\eq{LET} together with
the exact sum rule \eq{dispLET}
were used recently in \cite{kharzeev} and \cite{karsch} to obtain
information on the bulk viscosity
$\zeta=\frac1{18} \lim_{\omega{\to}0} \rho(\omega)/\omega$ near the QCD phase
transition.

It is not our goal here to discuss the equality of the left-hand
sides of \eqs{dispLET} and (\ref{LET})
nor possible contact
terms to be added to the right-hand side of \eq{dispLET};
this is discussed further by Romatschke and Son \cite{son}.
However, since
the results of \cite{kharzeev} and \cite{karsch} were based on the
assumption that \eq{dispLET}
is saturated by low $\omega$ 
(together with an Ansatz for the shape of the spectral function)
which is clearly in tension with the existence of the tail
\eq{opezeta},
we would like to investigate the importance of this tail.

Let us thus try to estimate the contribution from the
ultraviolet region $\omega\gsim \omega_\textrm{min}$ in the pure glue
theory.
Setting $\alphas(\omega)=2\pi/b_0 \log(\omega/\Lambda_\textrm{QCD})$
in \eq{opezeta} yields:
\bea
  \delta\GE^\zeta(0)_\textrm{UV} &\approx&
 \int_{\omega_\textrm{min}}^\infty \frac{d\omega}{\pi\omega}
\left[
  \frac{{-}C \pi T^\mu{}_\mu}{\log^2 \frac{\omega}{\Lambda_\textrm{QCD}}}
 - \frac{2\pi T^{00}}{\log^3 \frac{\omega}{\Lambda_\textrm{QCD}}}
\right]
\nl
&=&
C\frac{(3\PP{-}\EE)}{\log \frac{\omega_\textrm{min}}
      {\Lambda_\textrm{QCD}}}
-
 \frac{\EE}{\log^2 \frac{\omega_\textrm{min}}
      {\Lambda_\textrm{QCD}}}.
\label{uvbulk}
\eea
The logarithms in \eq{uvbulk} are never particularly large:
setting $\omega_\textrm{min}=2\pi T$ we estimate
$1/\log(\omega_\textrm{min}/\Lambda_\textrm{QCD}) \approx
 b_0\alphas(2\pi T)/2\pi\approx 0.4$
with $b_0=9$ in $\nF=3$ QCD and $\alphas=0.3$.
Thus we conclude that, at least in the pure glue theory,
whenever \eq{LET} is not parametrically large (e.g.,
$1/\cs^2$ large) compared to $0.4\EE$,
the sum rule \eq{dispLET} is very much sensitive to the
ultraviolet tail and
is not a clean probe of the $\omega\lsim \omega_\textrm{min}$ region.
It seems that this could affect the analysis of \cite{kharzeev}
and \cite{karsch}, at least
away from very close to the phase transition.

The weak-coupling limit of \eq{dispLET}
is particularly interesting: its left-hand side is of order
$g^6T^4$ ($g$ being evaluated at the scale $T$ from now on)
whereas its right-hand side receives a contribution
of order $g^4T^4$ from the $\omega\sim T$ region \cite{mooresaremi},
\be
 \int_{\omega\sim T} \frac{d\omega}{\pi\omega} \delta\rho(\omega)
\approx \int_0^\infty \frac{d\omega}{\pi\omega}
 \frac{\dA b_0^2\alphas^2
   \omega^4}{64\pi^3}n_\textrm{B}(\frac{\omega}{2})
= \frac{\dA b_0^2\alphas^2 T^4}{60}.
 \label{sumT}
\ee
At order $g^4T^4$ there is another contribution:
from the $\sim g^6T^4$ ultraviolet tail integrated
over a $\sim 1/g^2$ logarithmic range.
In the weak coupling limit this contribution is well-separated
and is given by \eq{uvbulk} evaluated at $\omega_\textrm{min}\sim T$.
The dominant term is the second term,
$-\EE b_0^2\alphas^2/4\pi^2$, which, remarkably, using \eq{loexp1}
is seen to exactly cancel \eq{sumT}.
The presently undetermined coefficient $C$ would only
become important at order $g^6T^4$.
Thus, the sum rule \eq{dispLET} is obeyed at order $g^4T^4$
but only when the ultraviolet tail is included.
This resolves a puzzle raised in \cite{mooresaremi}.

\subsection{Shear viscosity channel and discontinuity at $g^2\to 0$}

From \eq{opeeta}, the asymptotics of the thermal correction
$\delta\rho^\eta$
are proportional to $g^2$ times
an operator of strictly positive anomalous
dimension $\gamma$.
Upon resummation of logarithms its behavior will be proportional
to $(\log \omega)^{-1-a}$ (the $1$ coming from the running
of $g^2$ in an asymptotically free theory and $a>0$ coming from
the anomalous dimension): thus the unsubtracted dispersive
integral \eq{dispersion} converges.
As in the preceding subsection, higher order corrections
will not modify this result because the theory is asymptotically free.
The integrals should also converge in conformal theories such
as $\mathcal{N}=4$ super Yang-Mills, because in this case
all tails are associated with operators of strictly positive
anomalous dimensions $\gamma$, and decay like $\omega^{-\gamma}$.

Since the dispersive integral vanishes at $\omegaE\to\infty$
we can write in general:
\be
  \delta \GE^\eta(\omegaE) = \delta \GE^\eta(\infty)
+ \int_{-\infty}^\infty \frac{d\omega'}{2\pi(\omega'{-}i\omegaE)}
 \delta\rho^\eta(\omega')\,.
 \label{dispeta}
\ee
Note that convergence of the integral implies that $\delta\GE^\eta$
approaches a constant at infinity, so that $\delta\GE^\eta(\infty)$ is
well-defined.  According to \cite{son}, the left-hand side
at $\omegaE{=}0$
is determined
by hydrodynamical considerations.
In asymptotically free theories we believe
that the OPE coefficients of $\delta\GE^\eta(\infty)$
are saturated by a one-loop computation.%
 \footnote{
   Because other twist-two operators acquire positive anomalous
   dimensions, only three operators can appear in $\delta\GE^\eta(\infty)$:
   the traceless and trace part of the full stress tensor $T^{\mu\nu}$,
   and $\rmOm$.
   The coefficient of $T^\mu{}_\mu/g^2$
   vanishes at tree level \cite{novikovG} but a one-loop computation
   is needed to find that of $T^\mu{}_\mu$.
   The coefficients of twist-two operators and of $\rmOm$
   are determined at the tree level, but we believe a one-loop
   anomalous dimension matrix is necessary to carefully separate
   the total $T^{\mu\nu}$ from other twist-two operators, and $\rmOm$ from
   the trace $T^{\mu}{}_\mu$.
 }
Knowledge of both of these ingredients should yield interesting
exact sum rules, involving, at most,
the pressure, energy density and chiral condensates of QCD.
We hope to return to this question in the future.

Here we wish only to discuss a possible
discontinuity in the free theory limit $g^2\to 0$
of individual terms on the right-hand side of \eq{dispeta},
if the undetermined coefficient $D$ in \eq{opeeta} turns out to be nonzero.
Consider the term $T_g^{00}$ on the right-hand side
of the Euclidean OPE \eq{opeEta}.  At $g=0$ and $\omega=\infty$
it contributes a finite amount $T_g^{00}$ to
$\delta\GE(\infty)$, which discontinuously
changes to $(T_g^{00}+T_f^{00})/(1+n_\textrm{F}\TF/2\CF)$
at any small but finite coupling due to running, \eq{anom}.
On the other hand, at any finite but small coupling one has the
$\OO(g^2)$ tail \eq{opeeta} in the spectral function,
which is to be integrated over a $\OO(1/g^2)$ logarithmic
range similarly to the preceding subsection.
Its contribution is thus $\OO(1)$.
It is easy to convince oneself that it
exactly compensates for the discontinuity in $\GE(\infty)$.

Thus it might happen that
equations such as \eq{dispeta}
are only continuous at $g^2{=}0$ when both terms on the right-hand side
are included.
We hope to return in future work to shear channel sum rules in QCD
and in other theories, and to the question of whether
they actually contain strong ultraviolet tails.
In pure Yang-Mills, the coefficient $D$
is irrelevant and at the leading order such tails are absent.

\subsection{Strongly coupled $\mathcal{N}=4$ super Yang-Mills}
\label{sec:strong}

The operator spectrum of strongly coupled
multicolor ${\cal N}=4$ super Yang-Mills (SYM) has the very peculiar
property, that the only operators which do not develop
large anomalous dimensions 
$\sim \lambda^{1/4}$ are protected by supersymmetry
and have strictly vanishing anomalous dimensions \cite{adsreview},
where $\lambda=g^2N_\textrm{c}$ is the 't Hooft coupling.
There are no small nontrivial
anomalous dimensions in this theory, even at finite $\lambda$.

One way to find power corrections in spectral functions is,
as discussed in subsection \ref{sec:disp},
if the Euclidean OPE coefficients  have ``wrong'' phases.
This would certainly seem peculiar but we have no
general argument against this possibility.
However, the results of \cite{seiberg} suggest that
the OPE coefficients of protected operators in $\mathcal{N}=4$ SYM
are identical to those of the free theory,
for which this definitely does not happen.

Thus we will assume that power corrections 
to spectral functions at high frequencies
are associated with non-analytic terms in Euclidean
correlators.
Without anomalous dimensions the only sort of non-analyticity
allowed by the RGE (see subsection \ref{sec:RGE}) are
polynomial terms in momenta,
times single logarithms $\log(p^2/\mubar^2)$.
They lead to strictly polynomial terms in spectral functions.

By dimensional analysis and transversality
these are strictly forbidden in the spectral functions
of currents and stress tensors (except if they multiply the unit operator).
Thus these spectral functions are strictly
protected against medium-dependent power corrections.
It would be interesting to see whether polynomial corrections
actually occur in other spectral functions.
At finite but large $\lambda$, we expect
power tails $\sim \omega^{-n}$ with $n\sim \lambda^{1/4}$.

Thermal corrections to the spectral functions of R-currents
and stress tensors at strong coupling have been studied by Teaney
\cite{teaney06}
and observed, remarkably,
to decay exponentially fast at high energies.
This section generalizes his observation.


\begin{acknowledgments}

The author acknowledges useful discussions with G.~D.~Moore,
D.~T.~Son, P.~Romatschke
and K.~Dasgupta, and especially D.~T.~Son for finding
a mistake in the shear channel spectral function.
This work was supported by the Natural Sciences and
Engineering Research Council of Canada.

\end{acknowledgments}

\begin{appendix}

\section{Diagrammatic evaluation of $\rho^\mu_\mu$}
\label{app:calc}


This Appendix reproduces a calculation
of the trace $\rho^\mu_\mu(p)$
of the current spectral function,
using real-time Feynman diagrams.
We work in the high energy limit where $\rho=\Pi^>$
and drop all terms suppressed by
Boltzmann factors $\propto\exp(-p^0/2T)$, but
keep all power corrections.
The aim is to confirm the OPE result, \eq{operesult},
for $-\Pi^>{}^\mu_\mu=2\rho^T+\rho^L$.
For notational simplicity in this section we assume $\nF=1$.


\subsection{Outline of calculation}

\begin{figure}
\begin{center}
\includegraphics[width=0.48\textwidth]{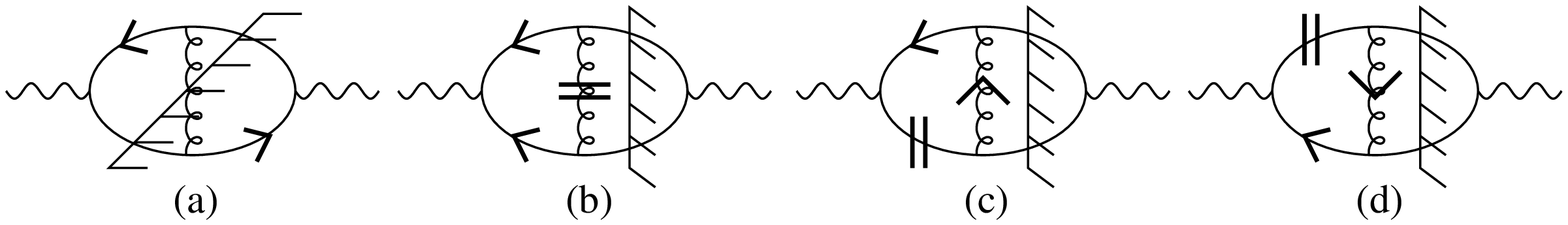}
\end{center}
\caption{Real-time diagrams of first topology contributing
to $\Pi^>$ (with the complex conjugate diagrams omitted).
The arrows show the time flow along
retarded propagators, not the
charge flow;  the doubly-dashed propagator is the fluctuation function
$G_{rr}$.
}
\label{fig:diags1}
\end{figure}

\begin{figure}
\begin{center}
\includegraphics[width=0.40\textwidth]{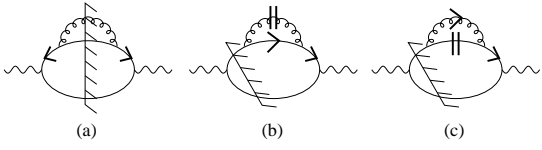}
\end{center}
\caption{Real-time diagrams of second topology contributing
to $\Pi^>$, in the notation of Fig.~\ref{fig:diags1}.
Not shown, the complex conjugates to (b)-(c).}
\label{fig:diags2}
\end{figure}

The two-loop diagrams contributing to the Wightman self-energy
$\Pi^>(q)$ are shown in figs.~\ref{fig:diags1} and
\ref{fig:diags2}.
The use the cutting rule of Weldon \cite{weldon},
which expresses the Wightman self-energy as a
product of two \emph{retarded} amplitudes
separated by Wightman (on-shell) propagators
(depicted as the main cut in the figures).
Its physical interpretation is as follows:
the main cut sums over intermediate states,
as is expected for a Wightman function,
and the amplitudes are retarded because intermediate states
are expanded in a basis of ``in'' states
(e.g., free theory states defined at $t\to -\infty$).

To evaluate the retarded amplitudes
on each side of the cut we use
the so-called Schwinger-Keldysh $(ra)$ formalism, as described
in \cite{keldyshreview}
 \footnote{
  Alternatively, these amplitudes are the
  analytic continuation of the Euclidean ones \cite{Evans}.
  Either way their evaluation at $n$-loop involves no more
  than $n$ statistical factors. This may be
  contrasted with the rules (of Kobes and Semenoff) employed
  in the second of reference \cite{dilepton} and \cite{gabellini}, in which
  terms in which all propagators carry statistical factors
  appear at intermediate steps (only to cancel out in the end).
}.
The resulting expressions are summarized
graphically in the figures: the arrowed
propagators are retarded propagators and the propagators with the double
cut are the fluctuation functions of this formalism
($G_{rr}$ propagators);
vertices are as in ordinary zero-temperature perturbation theory
(e.g., no complex conjugation appears).

\def\NB{\tilde{G}_{\textrm{B}}}
\def\NF{\tilde{G}_{\textrm{F}}}

Following the OPE philosophy we look for
propagators which can become soft, $q\sim T\ll p$.
Visual inspection reveals that no more than
one propagator can ever become soft simultaneously:
at least two hard particles must traverse the main cut,
and to channel their hard momenta to the external legs in all cases
requires at least two other hard propagators.
Thus we will organize the calculation around the propagator
which becomes soft.

At order $\alphas$ there is no need for HTL resummation
and the retarded propagators are
temperature-independent.
The temperature dependence is due to the statistical
factors entering the Wightman and $rr$ propagators,
\begin{subequations}
\begin{align}
G_R^{\mu\nu}(p)
   &= \frac{-i\delta^{\mu\nu}}{p^2+i\epsilon p^0},
&S_R(p)
   &=\frac{i\pslash}{p^2+i\epsilon p^0}, \\ \displaystyle
\label{ansatzcut}
\delta G^{\mu\nu}_{>,<,rr}(p)
  &= -\delta^{\mu\nu} \NB(p),
&\delta S_{<,>,rr}(p)
  &= -\pslash \NF(p),
\end{align}
\end{subequations}
with the vacuum cuts obeying
 $
  G_>(p)=2\textrm{Re}\, G_R(p)\theta(p^0)
 $.
We will not use the explicit forms
 $
  \tilde{G}_\textrm{B,F}(p)=2\pi\delta(p^2)n_\textrm{B,F}(|p^0|)
 $
until the end of the calculation;
up to then the sole purpose of the Ansatzes \eqs{ansatzcut}
is to simplify polarization sums.
Our metric signature is $({+}{-}{-}{-})$.


\subsection{Gluon condensate}

\def\vk{|\vec{k}|}

First we allow the gluon propagator in diagrams (a) and (b) of
Fig.~\ref{fig:diags1} to become soft.
Upon evaluating the Dirac trace this contribution may be written,
\begin{align}
 \label{agluondummy1}
 &\frac{{-}\Pi^>{}^\mu_\mu(p)}{g^2C_Fd_F}
  \supset
 32 \int_k \left[\NB(k)+\NB(-k)\right] \\
 & \hspace{1cm} \times \int_l 4\pi^2 \left(
 \frac{\delta(l^2)\delta(l_p^2)}{l_k^2 l_{pk}^2}  {+}
 \frac{\delta(l^2)\delta(l_{pk}^2)}{l_k^2 l_{p}^2} \right)
 l{\cdot}l_{pk} \,l_p{\cdot}l_k\,, 
 \nonumber
\end{align}
where
we have introduced the abbreviations $\int_l=\int \frac{d^4l}{(2\pi)^4}$,
$l_k=l{-}k$, $l_p=l{-}p$ and
$l_{pk}=l{-}p{-}k$, to be used in all what follows, and $k\ll l\sim p$
is the soft momentum.
It is kinematically impossible for two denominators
in \eq{agluondummy1}
to vanish simultaneously and only the real part (e.g., principal value)
of the propagators contributes.
The $l$-integration is Lorentz-covariant and becomes
elementary in the rest frame that is singled out by the $\delta$-functions.
Thus \eq{agluondummy1} yields:
\begin{align}
&
 \frac{1}{\pi}\int_k \NB(k)
  \left[
   2+\frac{(p^2{+}k^2)^2}{2p{\cdot k} \Delta}
  \ln\!\left( \frac{1{-}\left(\frac{p{\cdot}k{+}\Delta}{p^2}\right)^2}
    {1{-}\left(\frac{p{\cdot}k{-}\Delta}{p^2}\right)^2}\right)
\right.
\nl
  &\hspace{1cm} \left.
   +\frac{p^2{+}k^2}{\Delta} \ln\!\left(
  \frac{1{+}\frac{   p{\cdot}k{+}\Delta}{p^2}}
       {1{+}\frac{p{\cdot}k{-}\Delta}{p^2}}
  \frac{1{+}\frac{{-}p{\cdot}k{+}\Delta}{p^2}}
       {1{+}\frac{{-}p{\cdot}k{-}\Delta}{p^2}} \right)
  \right]\!,
 \label{agluon1}
\end{align}
with $\Delta= \sqrt{(p{\cdot k})^2-p^2k^2}$.

To evaluate the diagrams of the second topology, Fig.~\ref{fig:diags2},
without having to deal
with ill-defined expressions such as $\delta(l^2)/l^2$ (which would
appear in too literal an interpretation of diagrams (b)-(c)),
we write their sum as a discontinuity,
\begin{align}
\frac{{-}\Pi^>{}^\mu_{\mu}(p)}{g^2C_Fd_F} &\supset
16 \int_k \left[\NB(k)+\NB(-k) \right]
 \int_l 2\pi\delta(l_p^2)
 \nl
 & \hspace{1cm}
\times 2\textrm{Im}\,
\frac{2l{\cdot}l_p \,l{\cdot}l_k -l^2l_p{\cdot}l_k}
{(l^2+i\epsilon l^0)^2 (l_k^2+i\epsilon l_k^0)}\,,
\label{agluondummy2}
\end{align}
where we have also included the contribution with the self-energy
inserted on the lower propagator.
In our kinematic regime the poles of the denominators
are disjoint and occur at positive energies, $l^0>0, l_k^0>0$.
The discontinuity across the squared propagator $1/(l^2)^2$
may be conveniently evaluated by integration by
parts along $l_p^\mu\partial_{p^\mu}$, yielding
the formula%
 \footnote{
  An alternative
  way of deriving this result is to treat the self-energy
  insertion as a correction to the external states,
  in which case at this order one gets
  thermal mass shifts and wave-function renormalization factors.
  See, for instance, \cite{gabellini}.
 }:
\begin{align}
2\,\mbox{Im} \int_l \frac{\delta(l_p^2) F(l)}{(l^2{+}i\epsilon l^0)^2}
&=
\frac{1}{p^2} \int_l \delta(l_p^2) 2\pi\delta(l^2) \left(1{+}l_p^\mu
\frac{\partial}{\partial l^\mu}\right) \!F(l)\,,
\label{usefulse}
\end{align}
%
%
for $F(l)$ any function of $l$ regular at $l^2=0$.

The total imaginary part in \eq{agluondummy2} is the sum of that
from \eq{usefulse}
and from that across the $1/l_k^2$ propagator,
$2\Im 1/(l_k^2+i\epsilon l_k^0)=-2\pi\delta(l_k^2)$.
Upon performing the $l$ integration we obtain
\begin{align}
\hspace{-0.2cm}
& \frac{1}{\pi} \int_k
\NB(k) \left[
{-}4+\frac{p{\cdot}k}{\Delta}
 \ln\!\left( \frac{1{-}\left(\frac{p{\cdot}k{+}\Delta}{p^2}\right)^2}
    {1{-}\left(\frac{p{\cdot}k{-}\Delta}{p^2}\right)^2}\right)
\right]\!.
\label{agluon2}
\end{align}

Our final result for the sensitivity to the gluon distribution in
the medium is the sum of \eqs{agluon1} and (\ref{agluon2}).

\subsection{Fermion condensate}

Letting the lower fermion propagator become soft
in Fig.~\ref{fig:diags1} (a) and in its left-right flip,
or the $rr$ propagators with similar positions in (c) and its conjugate,
gives a contribution:
\begin{align}
 &\frac{{-}\Pi^>{}^\mu_{\mu}(p)}{g^2C_Fd_F}
 \!\supset
  {-}64\!\int_k \NF(k) \! \int_l 4\pi^2\!\!\left[
  \frac{\delta(l^2)\delta(l_{pk}^2)}{l_p^2} {+}
  \frac{\delta(l^2)\delta(l_{p}^2)}{l_{pk}^2}\right]
  \nl & \hspace{2cm}
 \times
       \frac{l_p{\cdot}(p{+}k) \,l{\cdot}k}{(p{+}k)^2}
\nl
& =\! 
  \frac{1}{\pi} \!\int_k \!\NF(k)\!\left[
\frac{2p\cdot k}{(p{+}k)^2} {-}
   \frac{(p^2{-}k^2)(2k{\cdot}p{+}k^2)}{(p{+}k)^2\Delta}
    \ln \frac{1{+}\frac{p{\cdot}k{+}\Delta}{p^2}}
             {1{+}\frac{p{\cdot}k{-}\Delta}{p^2}}
  \right]\!\!,
\label{aferm1}
\end{align}
where our notation and techniques are as in the previous subsection.
Contributions in which
the upper fermion propagators are allowed to become
soft are similar, but with $k$
replaced by $-k$; these will have to be included at the end.

The diagram of Fig.~\ref{fig:diags2}
receives a contribution from when
the upper fermion propagator becomes soft,
\begin{align}
\frac{{-}\Pi^>{}^\mu_{\mu}(p)}{g^2C_Fd_F} \supset
{-}32\,{\rm \Im}\!\!\int_k \NF(k) \!\int_l
\frac{2\pi \delta(l_p^2)
   (l^2 l_p{\cdot}k - 2l{\cdot k}\,l{\cdot}l_p)}
 {(l^2+i\epsilon l^0)^2(l_k^2+i\epsilon l_k^0)}
\nl
=
 \frac{1}{\pi} \int_k \NF(k)\left[
-1 + \frac{{-}\frac{1}2 p^2+p{\cdot k}}{\Delta} \ln\left(
 \frac{1-\frac{p{\cdot}k{+}\Delta}{p^2}}
      {1-\frac{p{\cdot}k{-}\Delta}{p^2}} \right)
\right],
\label{aferm2}
\end{align}
and from when the lower fermion
propagator becomes soft,
\bea
\frac{{-}\Pi^>{}^\mu_{\mu}(p)}{g^2C_Fd_F} &\supset&
16 \int_k \NF(k) \int_l \delta(l^2)\delta(l_{pk}^2)
\frac{ k{\cdot}(p{+}k) - l{\cdot}k}{(p{+}k)^2} \nl
&=& \frac{1}{\pi} \int_k \NF(k) \left[
  \frac{k{\cdot}(p{+}k)}{(p{+}k)^2} \right].
\label{aferm3}
\eea

The total sensitivity of $\Pi^>{}^\mu_\mu$ to fermions moving
in the medium is the sum of Eqs.~(\ref{aferm1}),
(\ref{aferm2}) and (\ref{aferm3}) and the
same objects with $(k\to{-}k)$.

\subsection{Expansion in $1/p$}

Each of the contributions Eqs. (\ref{agluon1}), (\ref{agluon2})
(\ref{aferm1}), (\ref{aferm2}) and (\ref{aferm3})
is free of infrared divergences and is moreover \emph{local} in $k$.
That is, each admits a Taylor expansion in positive powers of
$p{\cdot}k/p^2$ and $\Delta^2$, as is readily verified
from their parity under $\Delta\to{-}\Delta$.

This is the main point of this analysis: divergences and non-localities
have cancelled out in the sum over cuts, for each individual diagram.
A Minkowski-space OPE thus works for each diagram.
Furthermore,
upon summing the diagrams, we find that the term of order $(p^2)^0$
cancels in the $1/p^2$ expansion of the sum of
Eqs.~(\ref{agluon1}) and 
     (\ref{agluon2})
as required by gauge invariance, since this would correspond
to a non-invariant $A_\mu A^\mu$ condensate: gauge invariance
upon summing diagrams.

The leading nontrivial term in the expansion arise at order
$1/p^2$,
\begin{align}
\frac{{-}\Pi^>{}^\mu_\mu(p)}{g^2C_Fd_F} &\approx
 \frac{1}{\pi p^2}
 \int_k \left[
 \NF(k) \frac{8(p{\cdot}k)^2+ \frac{8}{3} \Delta^2}{p^2} \right.
\nl
 &\hspace{1.5cm}\left.
  -\NB(k) \frac{2(p{\cdot}k)^2 + \frac23 \Delta^2}{p^2}
+\rmO(\frac{k^4}{p^2}) \right]
\nl
&\simeq \frac{4p_{\mu}p_\nu}{3\pi p^4} \left[
  \frac{2}{d_\textrm{F}} \langle T^{\mu\nu}_f\rangle
 -\frac{1}{d_\textrm{A}} \langle T^{\mu\nu}_g\rangle\right]\!,
\label{directresg}
\end{align}
in complete agreement with the
OPE result for $(2\rho^T{+}\rho^L)(p)$ when $\nF=1$,
\eqs{operesult}.

\end{appendix}

\end{document}